\documentclass[12pt]{article}

\usepackage{scicite}
\usepackage{ulem} 
\usepackage{graphicx}
\usepackage{amsmath}
\usepackage{amsfonts}
\usepackage{color}
\usepackage{ulem}
\usepackage[normalsize,bf]{caption}
\usepackage{a4wide} 

\usepackage{times}

\newenvironment{sciabstract}{%
\begin{quote} \bf}
{\end{quote}}

\newcounter{lastnote}

\title{Sample and population exponents of generalized Taylor's law}

\author{Andrea Giometto$^{1,2\ast}$, Marco Formentin$^{3,4\ast}$, Andrea Rinaldo$^{1,5}$,\\ Joel E. Cohen$^6$,  Amos Maritan$^3$\\
\small{$^1$Laboratory of Ecohydrology (ECHO/IIE/ENAC), EPFL, CH-1015 Lausanne, Switzerland.}\\
\small{$^2$Department of Aquatic Ecology, Eawag, CH-8600 D\"ubendorf, Switzerland.}\\
\small{$^3$Dipartimento di Fisica ed Astronomia, Universit\`a di Padova, I-35131 Padova, Italy.}\\
\small{$^4$UTIA, Academy of Sciences of the Czech Republic, CZ-18208 Prague, Czech Republic.}\\
\small{$^5$Dipartimento ICEA, Universit\`a di Padova, I-35131 Padova, Italy.}\\
\small{$^6$Laboratory of Populations, The Rockefeller University and Columbia University,}\\ \normalsize{New York, NY 10065-6399 USA.}
\\
\small{$^\ast$ To whom correspondence should be addressed; E-mail: andrea.giometto@epfl.ch,}\\ 
\small{marco.formentin@ruhr-uni-bochum.de}
}

\date{}

\begin{document}

\baselineskip23pt

\maketitle

\begin{sciabstract}
Taylor's law (TL) states that the variance $V$ of a non-negative random variable is a power function of its mean $M$, i.e. $V=a M^b$. The ubiquitous empirical verification of TL, typically displaying sample exponents $b \simeq 2$, suggests a context-independent mechanism. However, theoretical studies of population dynamics predict a broad range of values of $b$. Here, we explain this apparent contradiction by using large deviations theory to derive a generalized TL in terms of sample and populations exponents $b_{jk}$ for the scaling of the $k$-th vs the $j$-th cumulant (conventional TL is recovered for $b=b_{12}$), with the sample exponent found to depend predictably on the number of observed samples. Thus, for finite numbers of observations one observes sample exponents $b_{jk}\simeq k/j$ (thus $b\simeq2$) independently of population exponents.  Empirical analyses on two datasets support our theoretical results.
\end{sciabstract}

\newpage

Taylor's law (TL) \cite{taylor61}, also known as fluctuations scaling in physics, is one of the most verified patterns in both the biological \cite{ramsayer12,cohen13b,giometto13,fronczak10,banavar07} and physical \cite{eisler08} sciences. TL states that the variance of a non-negative random variable $V=\mbox{Var}[X]$, is approximately related to its mean $M=\mathbb{E}[X]$ by a power law, that is, $\mbox{Var}[X]=a \mathbb{E}[X]^b$, with $a>0$ and $b\in\mathbb{R}$. In ecology, the random variable of interest is generally the size or density $N$ of a censused population and TL can arise in time (i.e., the statistics of $N$ are computed over time) or in space (i.e., the statistics are computed over space). The widespread verification of TL has led many authors to suggest the existence of a universal mechanism for its emergence, although there is currently no consensus on what such mechanism would be. Various approaches have been employed in the attempt, ranging from the study of probability distributions compatible with the law \cite{tweedie46,jorgensen87,kendal11a} to phenomenological and mechanistic models \cite{hanski87,keeling00,kilpatrick03,kalyuzhny14}. Although most empirical studies on spatial TL report an observed sample exponent $b$ in the range $1$--$2$ \cite{taylor61,anderson82}, mostly around $b\simeq2$ \cite{anderson82} (see also Fig. 10(g) in \cite{taylor82}), population growth models \cite{cohen13a,cohen13b,cohen14a,cohen14b,jiang14} can generate TL with any real value of the exponent. Moreover, theoretical investigations of multiplicative growth models in correlated Markovian environments \cite{cohen14a,cohen14b} have shown that the exponent $b$ can undergo abrupt transitions following smooth changes in the environmental autocorrelation.

Here, we distinguish between values of $b$ derived from empirical fitting (sample exponents) and values obtained via theoretical models that pertain to the probability distribution of the random variable $N$ (population exponents). We show that in a broad class of multiplicative growth models, the sample and population exponents coincide only if the number of observed samples or replicates is greater than an exponential function of the duration of observation. Among the relevant consequences, we demonstrate that the sample TL exponent robustly settles on $b\simeq 2$ for any Markovian environment observed for a duration that is larger than a logarithmic function of the number of replicates. Accordingly, abrupt transitions in the sample TL exponent can only be observed within relatively short time windows when the number of observations is limited.

Let us consider multiplicative growth models in Markovian environments \cite{cohen14a,cohen14b}. Let $N(t)$ be the density of a population at time $t$ and assume that the initial density is $N_0>0$. $N(t)$ is assumed to undergo a multiplicative growth process such that: 
\begin{equation}
N(t) \ = \ N_0 \ \prod_{n=1}^t A_n.
\label{multiplicativeN}
\end{equation}
The values of the multiplicative growth factors $A_i$ are determined via a finite-state homogeneous Markov chain with state space $\chi=\{r,s\}$ (we assume, without loss of generality, $r>s$ and $N_0=1$) and transition matrix $\Pi$ with $\Pi(i,j)>0$ for all $i,j \in \chi$ (see Supplementary Materials). In our notation, $\Pi(i,j)$ is the one-step probability to go from state $i$ to state $j$, i.e., $\Pi(i,j)=\mbox{Prob}(A_{n+1}=j|A_n=i)$. For the sake of clarity, we will restrict our discussion to symmetric transition matrices, with $\Pi(i,j)=\lambda$ for $i \neq j$. We derive (see Supplementary Materials) exact results on both sample and population TL exponents for a broad class of multiplicative processes, including state-spaces with size higher than $2$ and non-symmetric transition matrices.

By adopting large deviation theory techniques \cite{denhollander08,touchette09} and finite sample size arguments \cite{redner90}, we show \cite{derivazione} that for any choice of $\Pi$ and $\chi$, the sample mean and variance in a finite set of $R$ independent realizations of the process obey TL asymptotically as $t \to \infty$ with exponent $b \simeq 2$, even if the population mean and variance of $N(t)$ satisfy TL with exponent $b \neq 2$. Our analysis reveals two regimes ($t \gg \log R$ and $1 \ll t \ll \log R$, respectively \cite{logR}) where the sample TL holds with different exponents. In the former regime, sample exponents inevitably tend to $b \simeq 2$ independently of model specifications. In the latter, sample exponents accurately approximate population ones, which can be computed analytically and may differ from $b = 2$. Fig. 1 shows that simulation results and theoretical predictions in the two regimes are in excellent agreement. Fig. 2 shows the temporal evolution of the sample TL exponent, which crosses over from the approximate value of the population exponent (Eq. S6) at small times to the asymptotic prediction $b \simeq 2$ at larger times (Eq. S10).

We derive a generalized TL that involves the scaling of the $k$-th moment vs the $j$-th moment of the distribution of $N(t)$. Exact results \cite{derivazione} show that the generalized TL
\begin{equation}
\mathbb{E}[N^k(t)] = a_{jk} \mathbb{E}[N^j(t)]^{b_{jk}}
\end{equation}
holds asymptotically in $t$ for any choice of $j$ and $k$ (including non-integer values), both for population and sample moments (the positivity of $\Pi$ ensures that the same scaling relationship holds between the $k$-th and $j$-th cumulants, see Supplementary Materials). In accordance with the above results on the conventional TL (recovered in this framework with the choice $j=1$, $k=2$), two regimes exist: if $1 \ll t \ll \log R$, sample moments and cumulants accurately approximate population ones (and the value of $b_{jk}$ can be computed analytically); if $t \gg \log R$, the generalized TL exponent approximates $b_{jk}\simeq k/j$ (Fig. S4\textbf{C-D}).

In ecological contexts, the number of realizations $R$ that determine the possible convergence of sample and population TL exponents could refer, for instance, to independent patches experiencing different realizations of the same climate \cite{cohen14a}. In an established ecosystem, species have been present for several generations, and one might assume that the system is in the asymptotic regime $t \gg \log R$. Within this perspective, we tested the prediction that for large $t$ sample exponents satisfy the relation $b_{jk}=k/j$ (including the conventional TL) on two datasets.

A first example is drawn from a long-term census of plots within the Black Rock Forest (BRF) \cite{cohen13b}. It was shown that the Lewontin-Cohen model describes the population dynamics of trees in BRF and provides an interpretation of the TL exponent \cite{cohen13b}. Here, we computed, for each year $t$, the spatial sample moments $\langle N^k \rangle(t)$ of tree abundance across plots and we found that the least-squares slopes $b_{jk}$ of $\log \langle N^k(t) \rangle$ versus $\log \langle N^j(t) \rangle$ (Table S1) are compatible with the asymptotic model prediction $b_{jk}=k/j$ (see Supplementary Materials).

A second example uses the data collected by P Den Boer \cite{denboer77}, who measured abundances of carabid beetles in various sites across the Netherlands within a $200$-km$^2$ area for $8$ consecutive years. The dataset was shown to support the conventional spatial TL \cite{hanski87}. We computed the sample moments of carabid beetles abundance, $\langle N^k(t) \rangle$, across similar sites (either woodland or heath), for each species separately and year $t$. In the intra-specific analysis (Fig. \ref{jk}), linear regressions of  $\log \langle N^k(t) \rangle$ vs $\log \langle N^j (t) \rangle$ for $t=1,\dots,Y$ ($Y$ is the total number of years) gave the estimate of the sample exponent $b_{jk}$ for each species. Frequency histograms of empirical exponents $b_{jk}$ are shown in Fig. S5 (see an example in the lower inset of Fig. \ref{jk}). A one-sample $t$-test does not reject the null hypothesis that the sample mean of $b_{jk}$ does not differ significantly from the theoretically predicted mean $k/j$ (see Fig. S5). In the inter-specific analysis (Fig. \ref{jk_interspecific}), we calculated the least-squares slope $b_{jk}$ (for $j=1$) of $\log \langle N^k \rangle$ versus $\log \langle N \rangle$ across all species at a given year and site type (Tables S2, S3). The empirical exponents $b_{jk}$ for all years are compatible with the asymptotic model prediction $b_{jk}=k/j$, as are the mean (across years and site type) exponents $b_{jk}$ (Table S4).

{This empirical confirmation and the novel finding that other demographic models predict the generalized TL with $b_{jk}=k/j$ (see Supplementary Materials) indicate that these predictions are probably insensitive to the details of dynamics, just as the original TL is quite robust \cite{fronczak10,kendal11a,xiao14}.}

In conclusion, we have uncovered a general mechanism that yields TL with the widely observed sample exponent $b \simeq 2$. {For a broad range of parameters within the class of multiplicative models, and other demographic processes, the generalized TL describes the scaling of moments and cumulants with the sample exponent $b_{jk}$ asymptotically equal to $k/j$. This phenomenon may be attributable to the finite size of both ecosystems and sampling efforts. TL may not reflect (or depend on) the underlying population dynamics. Our theoretical prediction is supported by two empirical examples and invites further testing. Notably, our study suggests that limited sampling efforts might hinder the observation of abrupt transitions in population exponents that were recently discovered for theoretical multiplicative growth processes. Because fluctuations in population abundances strongly affect ecological dynamics, in particular extinction risk, comparable real-world transitions may harm fish populations, forests and public health. Our calculation of the minimum number of samples required to observe such transitions may help to identify early signals of abrupt biotic change following smooth changes in the environment.}

\section*{Acknowledgements}
We thank Dr. Hugo Touchette for discussions and Dr. Markus Fischer for fruitful discussions and a careful reading of the manuscript. AR and AG acknowledge the support provided by the discretionary funds of Eawag: Swiss Federal Institute of Aquatic Science and Technology and by the Swiss National Science Foundation Project 200021\_157174.  MF has been partially supported by GA\v{C}R grant P201/12/2613. JEC acknowledges with thanks the support of U.S. National Science Foundation grant DMS-1225529 and the assistance of Priscilla K. Rogerson.

\newpage
\section*{Figures}
\begin{figure}[htbp]
\begin{center}
\includegraphics[width=0.6\columnwidth]{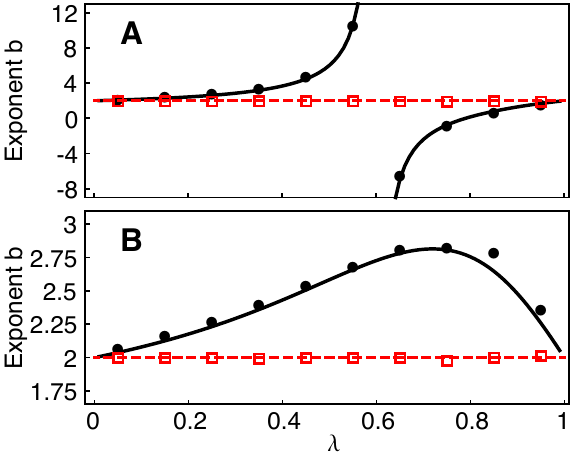}
\caption{TL exponent $b$ for different values of the transition probability $\lambda$. The sample exponent computed in simulations of a $2$-state multiplicative process with symmetric transition matrix in the two regimes $1 \ll t \ll \log R$ (black filled dots, $R=10^6$ up to time $t=10$) and $t \gg \log R$ (red open squares, $R=10^4$ up to time $t=400$) are in good agreement with predictions for the asymptotic population (black solid line, Eq. S6) and sample (red dashed line, $b=2$) exponents. In the simulations, the sample exponent $b$ was computed by least-squares fitting of $\log \mbox{Var}[N(t)]$ as a function of $\log \mathbb{E}[N(t)]$ for the last $6$ (black dots) and $200$ (red squares) time steps. In panel \textbf{(A)}, which reproduces \cite{cohen14a}, $\chi=\{r,s\}=\{2,1/4\}$ ($b$ displays a discontinuity, (see Supplementary Materials)); in panel \textbf{(B)} $\chi=\{r,s\}=\{4,1/2\}$ (in such a case, $b$ displays no discontinuity, (see Supplementary Materials)). Fig. S4 shows the generalized TL exponent $b_{23}$ in the same simulations.}
\label{b}
\end{center}
\end{figure}
\begin{figure}[p]
\begin{center}
\includegraphics[width=0.6\columnwidth]{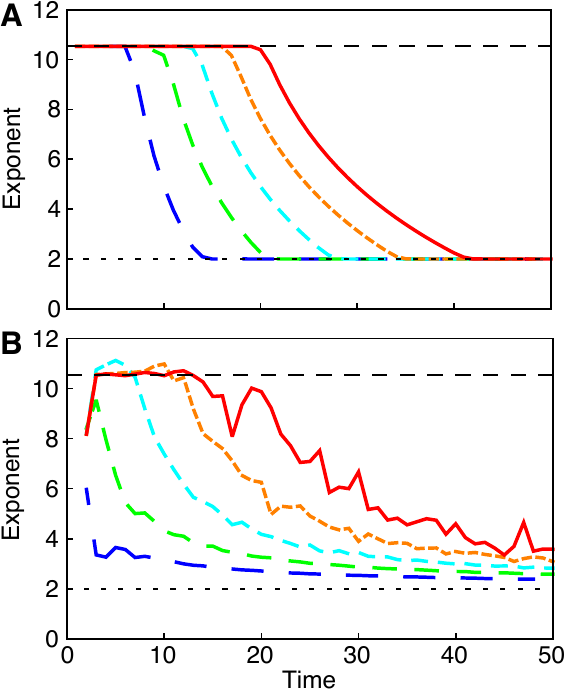}
\caption{Time evolution of the sample TL exponent. The sample exponent (computed as the slope of the curve $\log \mathbb{E}[N(t)^2]$ versus $\log \mathbb{E}[N(t)]$) crosses over from the approximate population exponent (Eq. S6, dashed upper horizontal line) at small times to $b\simeq2$ (dotted lower horizontal line) at larger times. The number of replicates $R=10^n$ increases exponentially from $10^2$ (blue dashed lines) to $10^6$ (red solid lines), while the crossover time increases approximately linearly. Here, $\chi=\{r,s\}=\{2,1/4\}$ and the transition probability in the symmetric $\Pi$ is $\lambda=0.55$. Panel \textbf{(A)} shows the theoretical prediction via Eq. S16. Panel \textbf{(B)} shows simulations results and the curves are averaged over $10^{8}/ R$ simulations (apart for the blue curve, which was averaged over $10^5$ simulations). Mismatches between panel \textbf{(A)} and \textbf{(B)} are due to the necessity to have $t$ and $R$ not too large to keep simulations feasible, while Eqs. S5, S6 and S10 hold true asymptotically in $t$.}
\label{b_decay}
\end{center}
\end{figure}
\begin{figure}[p]
\begin{center}
\includegraphics[width=0.6\columnwidth]{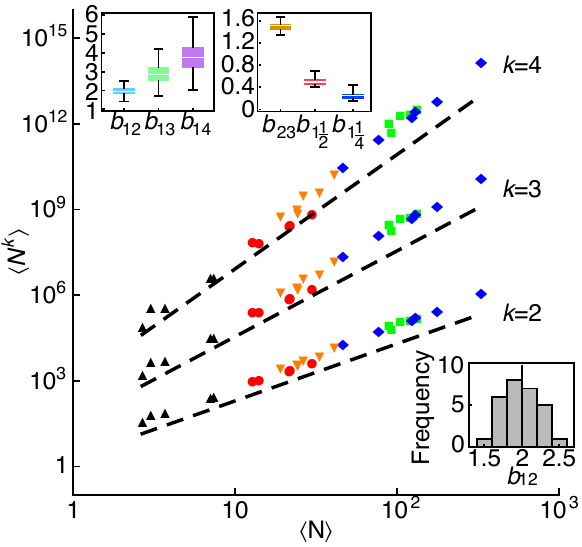}
\caption{Generalized TL for intra-specific patterns of carabid beetles abundance. Double logarithmic plot of $\langle N^k \rangle$ vs $\langle N \rangle$ for different species (identified by different colors and symbols), for consecutive years (each symbol refers to a single year $t$). For visual clarity, only $5$ species are shown. Dashed black lines of slopes $b_{1k}=k$ (asymptotic model prediction) are shown. Vertical offsets are introduced to aid comparison of slopes. Upper insets: box and whisker plots for the empirical distribution of intra-specific generalized TL exponents $b_{1k}$, showing the median (white horizontal line) and the $25\%$ and $75\%$ quantiles. Lower inset: histogram frequency distribution of the estimated conventional intra-specific TL exponent $b_{12}$. The vertical black line shows the value of the asymptotic predicted exponent $b=2$. See Supplementary Materials  for further details and statistical analysis.}
\label{jk}
\end{center}
\end{figure}
\begin{figure}[p]
\begin{center}
\includegraphics[width=0.6\columnwidth]{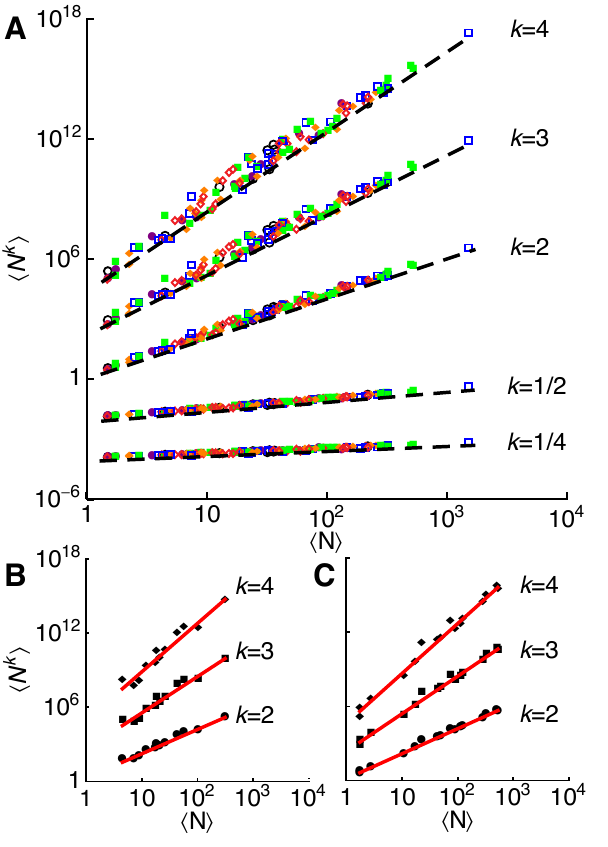}
\caption{Generalized TL for inter-specific patterns of abundance of carabid beetles. \textbf{(A)} Double logarithmic plot of $\langle N^k \rangle$ vs $\langle N \rangle$ for all species, years and site type. Each data point refers to a single species in one year and site type. The color and symbol code identifies data relative to the same year: 1961 (black open circles), 1962 (purple filled circles), 1963 (blue open squares), 1964 (green filled squares), 1965 (orange filled diamonds), 1966 (red open diamonds). Dashed black lines of slope $b_{1k}=k$ (asymptotic model prediction) are plotted next to the corresponding data series. Vertical offsets are introduced to aid comparison of slopes. \textbf{(B-C}) Examples of inter-specific moments scaling (each data point refers to a single species) for a single year and site type (\textbf{B}, woodland 1964 - \textbf{C}, heath 1964) used for the statistical analysis ((see Supplementary Materials), Tables S2, S3, S4, Fig. S6). The red lines are the least-squares regressions of $\log \langle N^k \rangle$ vs $\log \langle N \rangle$ across species.}
\label{jk_interspecific}
\end{center}
\end{figure}
\clearpage

\setcounter{equation}{0}
\setcounter{figure}{0}
\setcounter{table}{0}
\renewcommand{\theequation}{S\arabic{equation}}
\renewcommand{\thefigure}{S\arabic{figure}}
\renewcommand{\thetable}{S\arabic{table}}

\part*{Supplementary Materials}

\section{Methods}

\subsection{Theoretical Results}

We start from the multiplicative growth model in a Markovian environment introduced in \cite{cohen14a,cohen14b}, which includes as a special case (in the absence of autocorrelation) the Lewontin-Cohen model \cite{lewontin69,cohen13b}. Let $N(t)$ be the density of a population at time $t$ and assume that the initial density is $N_0>0$. $N(t)$ is assumed to undergo a multiplicative growth process such that: 
\begin{equation}
N(t) \ = \ N_0 \ \prod_{n=1}^t A_n.
\label{multiplicativeN}
\end{equation}
The values of the multiplicative growth factors $A_i$ are determined via a finite-state homogeneous Markov chain with state space $\chi=\{r,s\}$ (we label the state space $\chi=\{1\leftrightarrow r,2 \leftrightarrow s\}$ and, without loss of generality, assume $r>s$ and $N_0=1$) and transition matrix $\Pi$, with $\Pi(i,j)>0$ for all $i,j \in \chi$. In our notation, $\Pi(i,j)$ is the one-step probability to go from state $i$ to state $j$, i.e., $\Pi(i,j)=\mbox{Prob}(A_{n+1}=j|A_n=i)$. For the sake of clarity we will restrict this initial investigation to symmetric transition matrices, with $\Pi(i,j)=\lambda$ for $i \neq j$, but all results hold with minor changes also for non-symmetric matrices and in the case of a more general state space, as discussed in the subsequent paragraphs. The stationary distribution $\pi$ of the chain is unique and in the symmetric case satisfies $\pi(i)=1/2$, $i \in \chi$, for all $\lambda\in(0,1)$. We assume that the chain starts at equilibrium. 

We will show, under the broad assumptions just stated, that for any choice of $\Pi$ and $\chi$, the sample mean and variance in a finite set of $R$ independent realizations of the process (in an ecological example, e.g., $R$ sufficiently separated regions in space) obey TL asymptotically with exponent $b \simeq 2$, even when the population moments of the distribution of $N(t)$ satisfy TL with exponent $b \neq 2$. To distinguish between the two exponents, we will refer to the exponent of TL calculated with the sample mean and variance as the sample exponent and to that calculated via the population mean and variance as the population exponent. Correspondingly, we will distinguish between the sample and the population TL.

We introduce the empirical mean $L_t(z): \chi \to [0,1]$ defined as:
\begin{equation}
L_t(z)=\frac 1t \sum_{n=1}^t \delta_{A_n,z},
\end{equation}
where $\delta$ is the Kronecker's delta. The random measure $L_t(r)$ gives the fraction of times that $r$ appears in a realization of the Markov chain up to time $t$. $L_t$ satisfies a Large Deviation Principle (LDP) \cite{denhollander08} with rate function:
\begin{equation}
I_{\Pi}(x)=\sup_{u>0}\left[x\log\left(\frac{u_1}{(\Pi u)_1}\right)+(1-x)\log\left(\frac{u_2}{(\Pi u)_2}\right)\right]
\label{rate_function}
\end{equation}
where $x$ ($x \in [0,1]$) is the proportion of $r$ in a realization of the Markov chain up to time $t$ (correspondingly, the proportion of $s$ is $1-x$) and $u$ is a strictly positive vector in $\mathbb R^2$ (i.e., $u_1,u_2>0$). Stating that $L_t$ satisfies a LDP means that $\lim_{t \to \infty} \frac 1t \log \mathbb{P}(L_t(r) \in [x,x+dx]) =- I_{\Pi}(x)$. The rate function $I_{\Pi}(x)$ is convex ($d^2I_\Pi/dx^2>0$), attains its minimum at $x_{\min}=1/2$ with $I_\Pi(x_{\min})=0$ and is symmetric around $x_{\min}$ (Lemma IV.10 of \cite{denhollander08}, Theorems 3.1.2, 3.1.6 of \cite{dembo09}, Section 4.3 of \cite{touchette09}). The subscript $\Pi$ is used to indicate that the rate function depends on the transition matrix. Additionally, Eq. \ref{rate_function} depends on $u_1$ and $u_2$ only through $u\equiv u_2/u_1$; thus, by standard one-variable calculus, a long but explicit form of $I_{\Pi}(x)$ can be computed:
\begin{equation}
\begin{aligned}
I_\Pi(x) =& (x-1) \log \left[ 1-\lambda  \left(\frac{2 (\lambda -1) x}{\lambda +\sqrt{\lambda ^2+8 \lambda  (x-1) x-4 (x-1) x}-2 \lambda  x}+1\right) \right]-\\
&- x \log \left[ 1-\frac{\lambda  \left(\lambda +\sqrt{\lambda ^2+8 \lambda  (x-1) x-4 (x-1) x}-2 x\right)}{2 (\lambda -1) x}\right].
\label{ratef}
\end{aligned}
\end{equation}
The rate function does not depend on the values of the multiplicative factors $r$ and $s$. As in \cite{cohen14b}, we consider the ratio between $t^{-1}\log \mbox{Var} [ N(t) ]$ and  $t^{-1}\log \mathbb{E} [ N(t) ]$, but here we exploit the LDP, adopting Varadhan's lemma (Theorem III.13, \cite{denhollander08}), to perform such computation. First, since $\Pi$ is positive and $r\neq s$, it holds true that:
\begin{equation}\label{pos}
\lim_{t\rightarrow\infty}t^{-1}\log \mbox{Var} [ N(t) ]=\lim_{t\rightarrow\infty}t^{-1}\log \mathbb{E} [ N(t)^2 ].
\end{equation} 
See the Appendix in \cite{cohen14b} for a proof. Then, for the population moments of the population density $N(t)$, applying Varadhan's lemma, we have:
\begin{equation}
\lim_{t\to \infty} t^{-1} \log \mathbb{E}[N(t)^k]=\sup_{x\in[0,1]} \left[ k G(x)- I_{\Pi}(x) \right],
\label{varadhan}
\end{equation}
where $G(x)=x \log r + (1-x) \log s$. The population TL exponent $b$ (which depends on $\lambda$) can thus be computed as:
\begin{equation}
b(\lambda)=\frac{\sup_{x \in [0,1]}\left[ 2 G(x)-I_{\Pi}(x) \right]}{\sup_{x \in [0,1]} \left[ G(x) - I_{\Pi}(x) \right]}.
\label{exponent_infinite}
\end{equation}
For certain values of $r$ and $s$, $b(\lambda)$ can show a discontinuity at a critical value of the transition probability $\lambda$ (black line in Fig. 1\textbf{(A)}, see also Fig. \ref{b_rs}). The existence of such discontinuity was discovered and discussed in \cite{cohen14a}. An analysis of the critical transition probability is also available in section S.1.1.1. A generalized TL can be derived by adapting Eq. \ref{varadhan} to compute the scaling exponent for any pair of population moments as:
\begin{equation}
b_{jk}(\lambda)=\frac{\lim_{t\to \infty} t^{-1} \log \mathbb{E}[N(t)^k]}{\lim_{t\to \infty} t^{-1} \log \mathbb{E}[N(t)^j]}=\frac{\sup_{x\in[0,1]} \left[ k G(x)- I_{\Pi}(x) \right]}{\sup_{x\in[0,1]} \left[ j G(x)- I_{\Pi}(x) \right]}.
\label{exponent_infinite_jk}
\end{equation}
Discontinuities can also arise for these population exponents (section S.1.1.1). In the following, $b$ refers to the conventional TL population exponent ($b_R$ for the conventional TL sample exponent), while the generalized TL exponents are indicated with $b_{jk}$ (the distinction between sample and population exponents will be clear from the context). Both sample and population exponents were indicated as $b$ (or $b_{jk}$) in the main text to simplify the notation.
 
Eqs. \ref{exponent_infinite}, \ref{exponent_infinite_jk} hold true when one considers an infinite number of realizations of the multiplicative process, which ensures visiting the whole region $x \in [0,1]$. We now estimate the sample exponent, $b_R$, that is based on the sample mean and variance calculated over a finite set of $R$ realizations of the multiplicative process. We present first a heuristic derivation of the sample exponent. A more rigorous calculation of $b_R$ is given in the subsequent paragraph. We define $x_+$ as the value in $[0,1]$ such that the probability of a larger frequency $x$ of $r$ in $R$ runs of the Markov chain up to time $t$ is $1/R$ \cite{redner90}:
\begin{equation}
\mathbb P(L_t(r) \in (x_+,1])=\frac 1R.
\label{redner_R}
\end{equation}
With this definition, $x_+$ can be interpreted \cite{redner90} as the typical maximum frequency of $r$ in $R$ realizations of the chain. Analogously, we define $x_-$ as the value such that smaller values of the frequency of $r$ are observed with probability $1/R$, namely $\mathbb P(L_t(r) \in [0, x_-))=1/R$. For large $t$, one can adopt Varadhan's lemma (or Laplace's method of integration) to obtain, as a function of $t$, the approximate number of replicas $R$ needed to explore rare events (i.e., to compute $\mathbb P(L_t(r) \in (x_+,1])= R^{-1}$).
Approximately:
\begin{equation}
R \simeq \exp \left[ t I_{\Pi}(x_\pm) \right].
\label{exponential_R}
\end{equation}
Inversion of this formula (by taking the logarithm on both sides and expanding $I_\Pi$ in Taylor series around $x=x_{\min}$) gives $x_\pm\simeq \frac 12 \pm \sqrt{ \frac{1-\lambda}{2\lambda} \frac{\log R}t }$. Consequently, the sample TL exponent in an ensemble of $R$ realizations of the process can be approximated as:
\begin{equation}
b_R(\lambda,t) \simeq \frac{\sup_{x \in [x_-,x_+]}\left[ 2 G(x)-I_{\Pi}(x) \right]}{\sup_{x \in [x_-,x_+]} \left[ G(x) - I_{\Pi}(x) \right]},
\label{exponent_finite}
\end{equation}
where the dependence on $t$ is through $x_+$ and $x_-$. The zero of the rate function, $x_{\min}=1/2$, corresponds to the most probable value of the product in Eq. \ref{multiplicativeN}. Because $x_\pm\simeq \frac 12 \pm \sqrt{ \frac{1-\lambda}{2\lambda} \frac{\log R}t }$, for fixed $R$ the suprema in Eq. {\ref{exponent_finite} are computed over an increasingly narrower set around $x_{\min}$ (with $I_{\Pi}(x_{\min})=0$) as $t$ increases (Fig. \ref{figI_G}). Thus, for any finite number of realizations $R$, the sample exponent will approximate $\lim_{t\to\infty} b_R(\lambda,t) \simeq 2$ after a time $t^*$ that increases only logarithmically with $R$ (Eq. \ref{exponential_R} and Fig. 2), for any choice of $\lambda$, $r$ and $s$. For example, with $\lambda=0.5$, when $t=100$, in order to access to the extreme event $x_{+}=0.9$ (and $x_{-}=0.1$) one needs about $R\simeq 10^{16}$ replicates of the process.  Fig. 1 illustrates typical behaviors of sample and population exponents as a function of the transition probability $\lambda$ for the $2$-state multiplicative model with symmetric transition matrix. The black and red lines portray respectively the predicted asymptotic population and sample exponent (Eqs. \ref{exponent_infinite} and \ref{exponent_finite}), computed for different values of $\chi=\{r,s\}$ in the two panels. Dots and squares illustrate the sample exponents $b_R$ calculated via simulations in the regimes $t \ll \log R$ and $t \gg \log R$, respectively. Simulations results in the two regimes (dots and squares) and theoretical predictions (solid and dashed lines) show excellent agreement. Analogous considerations hold for the asymptotic sample exponent describing the scaling of the sample moments $\mathbb{E}[N(t)^k]$ with $\mathbb{E}[N(t)^j]$, which can be approximated as:
\begin{equation}
b_{jk}(\lambda,t) \simeq \frac{\sup_{x \in [x_-,x_+]}\left[ k G(x)-I_{\Pi}(x) \right]}{\sup_{x \in [x_-,x_+]} \left[ j G(x) - I_{\Pi}(x) \right]},
\label{exponent_finite_jk}
\end{equation}
which is the analogue of Eq. \ref{exponent_finite} for any pair of sample moments.

The above calculations identify the logarithmic dependence of $x_+$ on the number of realizations $R$, but rely on a number of approximations: the definition of $x_+$ (which, in a given realization, is a random variable), the computation of Laplace integrals (Eq. \ref{redner_R}) and the expansion of the rate function around $x_{\min}$ (Eq. \ref{exponential_R}). Such calculations can be made more rigorous if we consider the independent identically distributed random variables $X^i(t)=L_t^i(r)$, that is, $X^i(t)$ is the frequency of occurrence of the first state up to time $t$ in the $i$-th realization of the Markov chain ($i=1,\dots,R$). We now define $x^+=\max \{ X^1(t), \dots, X^R(t) \}$ and observe that:
\begin{equation}\label{inequality:1}
\frac{1}{t}\log \mathbb{P}(X^1(t)>x) \leq \frac{1}{t}\log \mathbb{P}(x^+>x) \leq \frac{1}{t}\log(R)+\frac{1}{t}\log \mathbb{P}(X^1(t)>x).
\end{equation}
For fixed $R$ (or, more generally, $\log R = o(t)$) and $x>1/2$, taking the limit ($\lim_{t \to \infty}$) in Eq. \ref{inequality:1} and knowing that $L^t(r)$ satisfies a LDP, one has:
\begin{equation}\label{equality:1}
\lim_{t\rightarrow\infty} \frac{1}{t}\log \mathbb{P}(x^+>x) =\sup_{y\in(x,1]} -I_{\Pi}(y)= -I_{\Pi}(x).
\end{equation}
Because $0<I_\Pi(x)\leq\infty$, Eq. \ref{equality:1} implies that $\lim_{t \to \infty} \mathbb{P}(x^+>x)=0$ for any $x>1/2$. An analogous calculation for $x^-=\min \{ X^1(t), \dots, X^R(t) \}$ shows that $\lim_{t \to \infty} \mathbb{P}(x^-<x)=0$ for any $x<1/2$. In this context, we can approximate the sample exponent at time $t$ with an analogue of Eq. \ref{exponent_finite}:
\begin{equation}
b_R(\lambda,t) \simeq \frac{\sup_{x \in [x^-,x^+]}\left[ 2 G(x)-I_{\Pi}(x) \right]}{\sup_{x \in [x^-,x^+]} \left[ G(x) - I_{\Pi}(x) \right]}.
\label{exponent_finite_P}
\end{equation}
In the narrow interval $[x^-,x^+]$ centered  around $x_{\min}$, $I_{\Pi}(x) \simeq 0$ and as a consequence $b_R(\lambda,t)~\simeq~2$ (Fig. \ref{figI_G}). More precisely, $|b_R(\lambda,t)-2|$ goes to $0$ in probability as $t$ tends to infinity. In fact, for every $\epsilon>0$, we have:
\begin{equation}\label{convergence:probability}
\mathbb{P} \left(|b_R(\lambda,t)-2|>\epsilon\right)\leq\mathbb{P}\left(x^+>\frac{1}{2}+\eta(\epsilon)\right)+\mathbb{P}\left(x^-<\frac{1}{2}-\eta(\epsilon)\right),
\end{equation}
where $\eta(\epsilon)$ is a function that goes to zero for $\epsilon\rightarrow 0$. Because of Eqs. \ref{inequality:1} and \ref{equality:1}, it follows that:
 \begin{equation}\label{convergence:probability:bis}
\lim_{t\rightarrow\infty}\mathbb{P} \left(|b_R(t)-2|>\epsilon\right)=0.
\end{equation}
Analogous considerations hold for the generalized TL describing the scaling of any pair of moments.

Eqs. \ref{exponent_finite} and \ref{exponent_finite_P} give the estimated sample exponent of TL asymptotically, ignoring the constant term in the scaling of the variance $V$ versus the mean $M$ as $\log V= b \log M + \log a$. For small $t$, $\log a$ can be of the same order of magnitude of $\log V$. Fig. 2 shows the crossover of the sample exponent (for fixed $R$, $\lambda$, $r$ and $s$) from the population exponent $b=b(\lambda)$ as in Eq. \ref{exponent_infinite} (observed when $t \ll \log R$) to $b \simeq 2$ (when $t \gg \log R$), where the sample exponent is calculated as the slope of the curve $\log \mathbb{E}[N(t)^2]$ versus $\log \mathbb{E}[N(t)]$ at time $t$ (thus not neglecting the constant term $\log a$). The sample moments are computed as:
\begin{equation}
t^{-1} \log \mathbb{E}[N(t)^k] \simeq \sup_{x\in[x_-,x_+]} \left[ k G(x)- I_{\Pi}(x) \right]
\label{smoments}
\end{equation}
(cf. Eq. \ref{varadhan}) in panel \textbf{(A)} and as the sample moments in simulations in panel \textbf{(B)}.

We now look at some generalizations of the stochastic multiplicative process considered above. The sample exponent in a finite set of $R$ independent realizations of the process is $b\simeq 2$ also for non-symmetric transition matrices $\Pi$. In the asymmetric case, the transition matrix is:
\begin{equation}
\Pi=\left(\begin{array}{cc}1-\lambda & \lambda  \\ \mu  & 1-\mu \end{array}\right),
\label{Piasym}
\end{equation}
with $0<\lambda,\mu<1$. The rate function $I_\Pi(x)$ is convex, attains its minimum at $x_{\min}=\pi(1)=\mu/(\lambda+\mu)$, where $\pi=(\pi(1),\pi(2)=\lambda/(\lambda+\mu))$ is the invariant measure for $\Pi$ and $I_\Pi(x_{\min})=0$. Only the value of the rate function at $x_{\min}$ and not the value of $x_{\min}$ is relevant for our argument. Due to asymmetries of $I_{\Pi}$, `left' (i.e., $x<x_-$) rare events could be easier to see than `right' (i.e, $x>x_+$) rare events or vice versa. In all cases, an exponentially large in $t$ number of replicates is needed to sample the tails with the correct weights. In this context, Eqs. \ref{exponent_infinite}, \ref{exponent_infinite_jk} and Eqs. \ref{exponent_finite}, \ref{exponent_finite_jk} are still valid and give, respectively, the asymptotic population and sample exponents.

{The previous considerations can also be extended to multiplicative processes $N(t)$ in more general Markovian environments with $w$ states and state space $\chi=\{r_1,\dots,r_w\}$, where all $r_i$ are strictly positive and at least two $r_i$ are different. We label the state space $\chi=\{1\leftrightarrow r_1, \dots, w \leftrightarrow r_w\}$. Let the transition matrix $\Pi$ be two-fold irreducible (i.e., $\Pi$ irreducible and $\Pi \ \Pi^\top$ irreducible, where $\Pi^\top$ is the transpose of $\Pi$). The rate function in Eq. \ref{rate_function} reads (Theorem IV.7 and Section IV.3 of \cite{denhollander08}, or Theorem 3.1.6 of \cite{dembo09}):
\begin{equation}
I_{\Pi}(\mu)=\sup_{u>0}\left[\sum_{v=1}^w\mu_v\log\frac{u_v}{(\Pi u)_v}\right],
\label{rate_function_general}
\end{equation}
where $u$ is a strictly positive vector in $\mathbb R^w$. Here, $\sum_{v=1}^w \mu_v=1$, and $\mu_v$ represents the proportion of $v$ after $t$ steps (for large $t$). The rate function is convex and $I_{\Pi}(\mu_{\min})=0$, with $\mu_{\min}$ the most probable state for large $t$ (Theorems 3.1.2, 3.1.6 of \cite{dembo09}, Section 4.3 of \cite{touchette09}). Eq. \ref{exponent_infinite}, with $x$ in the standard $w-1$ simplex in $\mathbb R^w$ and $G(x)=\sum_{i=1}^w x_i \log r_i$, gives the population scaling exponent of $\mathbb{E} [N(t)^2 ]$ with $\mathbb{E} [N(t) ]$. The two-fold irreducibility of $\Pi$ plus the condition that $r_i \ne r_j$ for some $i \ne j$ is the sharpest sufficient assumption that is presently known \cite{cohen14b} to guarantee that the limiting growth rate of the second moment equals the limiting growth rate of the variance; thus, Eq. \ref{exponent_infinite}, with $x$ in the standard $w-1$ simplex in $\mathbb R^w$ and $G(x)=\sum_{i=1}^w x_i \log r_i$, gives the population scaling exponent of $\mbox{Var} [N(t)]$ with $\mathbb{E} [N(t) ]$. Analogously, Eq. \ref{exponent_infinite_jk}, with $x$ in the standard $w-1$ simplex in $\mathbb R^w$ and $G(x)=\sum_{i=1}^w x_i \log r_i$, gives the population scaling exponent of $\mathbb{E} [N(t)^k ]$ with $\mathbb{E} [N(t) ]$. As far as the scaling of moments is of interest, the ergodicity (i.e., irreducibility and aperiodicity) of $\Pi$ (as opposed to the two-fold irreducibility) and $G(x)$ not identically equal to zero (which happens only if $r_i=1$ $\forall i$) are sufficient to compute the scaling exponents via Eqs. \ref{exponent_infinite}, \ref{exponent_infinite_jk}, modified as stated above. 
This is true because the ergodicity of $\Pi$ ensures that the empirical measure $L_t$ satisfies a LDP (Theorems 3.1.2 and 3.1.6 of \cite{dembo09}). Therefore, one can apply Varadhan's lemma (Theorem III.13, \cite{denhollander08}) to compute the limiting growth rate of the moments of $N(t)$ via Eq. \ref{varadhan}, with $x$ in the standard $w-1$ simplex in $\mathbb R^w$ and $G(x)=\sum_{i=1}^w x_i \log r_i$. The computation of the sample exponents $b_R$ and $b_{jk}$ is similar to that in the $2$-state case and the sample exponents approximate $b_R=2$ and $b_{jk}=k/j$ asymptotically in time; the proof is as follows. We consider the independent identically distributed random variables $Y^i(t)=|L_t^i-\mu_{\min}|$, where  $L_t^i=( L_t^i(r_1), \dots, L_t^i(r_w) )$ and the superscript $i$ indicates the $i$-th independent realization of the chain ($i=1,\dots,R$). We now define $y^+=\max \{ Y^1(t), \dots, Y^R(t) \}$ and observe that, for every $\epsilon>0$:
\begin{equation}\label{inequality:1w}
\mathbb{P}(y^+>\epsilon) \leq R \  \mathbb{P}(Y^1(t)>\epsilon).
\end{equation}
For fixed $R$ and $\epsilon$, taking the limit ($\lim_{t \to \infty}$) in Eq. \ref{inequality:1w} and knowing that $L_t^1$ satisfies a LDP (in particular, $\lim_{t\rightarrow\infty} \mathbb{P}(Y^1(t)>\epsilon) = 0$), one has:
\begin{equation}\label{equality:1w}
\lim_{t\rightarrow\infty} \mathbb{P}(y^+>\epsilon) = 0.
\end{equation}
In this context, we can approximate the sample exponent with:
\begin{equation}
b_R(\lambda,t) \simeq \frac{\sup_{|\mu-\mu_{\min}| < y^+}\left[ 2 G(\mu)-I_{\Pi}(\mu) \right]}{\sup_{|\mu-\mu_{\min}| < y^+} \left[ G(\mu) - I_{\Pi}(\mu) \right]}.
\label{exponent_finite_P}
\end{equation}
In the narrow region $|\mu-\mu_{\min}| < y^+$ centered around $\mu_{\min}$, $I_{\Pi}(\mu) \simeq 0$ and as a consequence $b_R(\lambda,t)~\simeq~2$. More precisely, $|b_R(\lambda,t)-2|$ goes to $0$ in probability as $t$ tends to infinity. In fact, for every $\delta>0$, we have:
\begin{equation}\label{convergence:probabilityw}
\mathbb{P} \left(|b_R(\lambda,t)-2|>\delta \right)\leq\mathbb{P}\left(y^+>\eta(\delta)\right),
\end{equation}
where $\eta(\delta)$ is a function that goes to zero for $\delta \rightarrow 0$. Because of Eq. \ref{equality:1w}, it follows that:
 \begin{equation}\label{convergence:probability:bisw}
\lim_{t\rightarrow\infty}\mathbb{P} \left(|b_R(t)-2|>\delta\right)=0.
\end{equation}
Analogous considerations hold for the generalized TL describing the scaling of any pair of moments. A standard saddle-point calculation suggests that the limiting growth rate of the variance is equal to the limiting growth rate of the second moment also for ergodic transition matrices, apart from peculiar cases (see \cite{cohen14b} for a discussion of a counterexample). The same argument suggests that the limiting growth rate of the $k$-th cumulant equals that of the $k$-th moment ($t^{-1} \log \mathbb{E}[N(t)^k]$) for large $t$. The suggested equivalence between the scaling exponents of cumulants and moments for ergodic $\Pi$ would allow extending the result on the sample TL ($b = 2$) and generalized TL ($b=k/j$) to the scaling of cumulants in $m$-step Markov chains, whose transition matrix is ergodic but not two-fold irreducible. However, pathological counterexamples may exist.} 

\subsubsection{Analysis of the discontinuity in $\lambda$ as a function of $r$ and $s$}
{A discontinuity in the population TL exponent $b$ (Fig. 1, Eq. \ref{exponent_infinite}) is present when the limiting growth rate of the mean abundance is zero, i.e., $\lim_{t\to \infty} \frac 1t \log\mathbb{E}[ N(t) ]=0$.  Let us consider Fig. \ref{figI_G} and fix $r$ and $s$ with $r\neq s$. The value of $\lambda$ shapes the form of $I_{\Pi}(x)$ (black curve in Fig. \ref{figI_G}); in particular, the second derivative can be easily calculated from Eq. \ref{ratef} and shown to increase for larger $\lambda$. A discontinuity may eventually appear for the value $\lambda=\lambda_c$ such that the curve $I_{\Pi}(x)$ and the line $G(x)$ (blue line in Fig. \ref{figI_G}) are tangent. In other words, $\lim_{t\to \infty} t^{-1}  \log \mathbb{E} [ N(t) ]=\sup_{x\in[0,1]} [G(x)-I_{\Pi}(x)]=0$ for $\lambda=\lambda_c$ such that
\begin{equation}\label{critical}
\log\frac{1}{2}\left[(1-\lambda_c)(r+s)+\sqrt{4(2\lambda_c-1)rs+(\lambda_c-1)^2(r+s)^2}\right]=0,
\end{equation}
with constraints $r,s>0$ and $0<\lambda_c<1$. $\lambda_c$ exists only for certain values of $r$ and $s$, thus a discontinuity in the population TL exponent $b$ is not always possible. Solving Eq. \ref{critical} with respect to $\lambda_c$ gives $\lambda_c=\frac{1-r-s+rs}{-r-s+2 rs}$; thus, for any given $s$, $\lambda_c=0$ for $r=1$ and $\lambda_c=1$ for $r=1/s$. For fixed $s \neq 1$ one has $d \lambda_c / dr >0$ (except for $r=s$ where $d \lambda_c / dr |_{r=s} =0$); thus, $\lambda_c$ exists for $0<r \leq 1/s$ and $r\geq 1$ if $s>1$ and for $1\leq r \leq 1/s$ if $s<1$ (see Fig. \ref{figure11}). Fig. \ref{phase} schematically illustrate the behavior of $b(\lambda)$ for different pairs $\{r,s\}$ of multiplicative factors. Discontinuities analogous to that of $b(\lambda)$ appear for certain values of $r$, $s$ and $\lambda$ in the population exponents $b_{jk}$ (Eq. \ref{exponent_infinite_jk}), when  $\lim_{t\to \infty} t^{-1} \log \mathbb{E} [ N(t)^j ]=\sup_{x\in[0,1]} [jG(x)-I_{\Pi}(x)]=0$.\\

\subsubsection{Compatibility of Eq. \ref{exponent_infinite} here and Eq. 8 in reference \cite{cohen14b}}
We show here that Eq. \ref{exponent_infinite} coincides with Eq. 8 in reference \cite{cohen14b}, under the assumption (stronger than in \cite{cohen14b}) that the transition matrix $\Pi$ is positive and $r \neq s$. The rate function Eq. \ref{rate_function} can be written as (Section 4.3 of \cite{touchette09} or Theorem 3.1.7 of \cite{dembo09}) $I_{\Pi}(x)=\sup_{q}\left\{ qx - \log \zeta(\Pi_q) \right\}$, where $\Pi_q$ is the matrix with elements $\Pi_q(i,j)=\Pi(i,j) \exp(q \delta_{j,1})$, and $\zeta(\cdot)$ indicates the spectral radius (i.e., the Perron-Frobenius eigenvalue). $\zeta(\Pi_q)$ is unique and analytic in $q$; thus, $\xi(q) \equiv \log \zeta(\Pi_q)$ is differentiable and the rate function can be expressed as $I_\Pi(x)=q(x)x - \xi(q(x))$, where $q(x)$ is the unique solution of $\xi'(q)=x$. Eq. \ref{varadhan} for the $k$th moment of $N(t)$ then reads $\lim_{t\to \infty} \frac 1t \log \mathbb{E}[N(t)^k]=\sup_{x\in[0,1]} \left[ k G(x) - q(x)x + \xi(q(x)) \right]$. The argument of the supremum is maximum at $x^*$ such that $k \log (r/s) - q(x^*) = 0$, that is, $x^*=\xi' \left(k \log (r/s) \right)$. Thus, evaluating the supremum one has $\lim_{t\to \infty} \frac 1t \log \mathbb{E}[N(t)^k]=k \log s + \xi(k \log (r/s))=\log \left[ s^k \zeta\left(\Pi_{k \log(r/s)} \right) \right]=\log \zeta(\Pi \ \mbox{diag}(r,s)^k)$, which coincides with Eqs. 13, 14 of reference \cite{cohen14b} (Eqs. 13, 14 in reference \cite{cohen14b} are expressed in terms of the column-stochastic matrix $\Pi^\top$ that corresponds to the row-stochastic matrix $\Pi$; because $\zeta(\mbox{diag}(r,s)^k \ \Pi^\top)=\zeta(\Pi \ \mbox{diag}(r,s)^k)$, the equations coincide). The compatibility of Eq. \ref{exponent_infinite} here with Eq. 8 in \cite{cohen14b} follows directly.

\subsection{Software and numerical analysis}
All analyses performed in this study were done with the software \textit{Wolfram Mathematica}, version 9.0. Simulation of the multiplicative process in Eq. \ref{multiplicativeN} in software with finite precision is subject to numerical underflow and overflow. This may result in errors in the estimation of exponentially growing or declining abundances after very few steps, if simulations are not performed carefully. For example, \textit{MathWorks Matlab} cannot represent the numbers $4^{1000}$ and $(1/4)^{1000}$, which are evaluated respectively as $+\infty$ and $0$. Additionally, finite-precision software might misrepresent the sum of very large and very small numbers. By using rationals and integers, \textit{Wolfram Mathematica} allows infinite precision calculations and thus simulates correctly the multiplicative process in Eq. \ref{multiplicativeN} and computes exactly the moments at every time $t$. Therefore, all numerical calculations in this study are free of underflow and overflow issues.

\subsection{Generalized TL for tree abundance in the Black Rock Forest (USA)}

We tested the predictions of the multiplicative growth model by using a dataset of tree abundance from six long-term plots in the Black Rock Forest (BRF), Cornwall, NY, USA. Therein, it was shown \cite{cohen13b} that the Lewontin-Cohen (LC) model (a particular case of the multiplicative model studied here) described the population dynamics of trees in the BRF. The interpretation of the six plots as distinct and independent replicates of the LC model is supported by statistical analysis \cite{cohen13b} and allowed relating the model predictions to the spatial TL. Here, we used the same dataset to show that the generalized TL holds with sample exponent $b_{jk}=k/j$. We computed the moments ratios $\langle [N(t)/N_0]^k \rangle$, where the symbol $\langle \cdot \rangle$ identifies the sample mean across the six plots of BRF and $N_0$ is the number of trees at the start of the census in $1931$. Following \cite{cohen13b}, we tested whether the moments of the spatial density ratio $N(t)/N_0$ in the five most recent censuses satisfied TL and the generalized TL with $b_{jk}=k/j$. Table \ref{table_BRF} reports the slopes of the least-squares linear regressions of $\langle [N(t)/N_0]^k \rangle$ versus $\langle [N(t)/N_0]^j \rangle$, which are all compatible with the model prediction $b_{jk}=k/j$. The BRF dataset thus provides an empirical example where the multiplicative model satisfactorily describes the underlying dynamics and the generalized TL holds asymptotically as the model predicts.

\subsection{Generalized TL for carabid beetles abundance}

\subsubsection{Intra-specific TL data analysis}
The intra-specific form of TL and the generalized scaling relationship between higher moments (Eq. 2) were tested using abundance data from $26$ species of carabid beetles. We have limited the analysis of the intra-specific TL to the set of species that were present in all sites in each given year.  We have followed the researchers who collected the carabid beetles abundance data \cite{denboer77} in excluding species with year-samples with zero individuals in at least one of the sites from the statistical analysis. In fact, the authors of \cite{denboer77} declared that they were unable to differentiate sites where a species was not present from sites where the density of such species was so low that no catches were realized. For each species, we selected data from a minimum of $3$ to a maximum of $6$ sites (all either woodland or heath, see \cite{denboer77}) and from a minimum of $4$ to a maximum of $6$ consecutive years. The precise number of sites and years varied for each species, depending on the number of sites and years in which at least one individual of such species was found in each site. The moments of species abundance were calculated separately for each species and for each available year. Linear regressions of  $\log \langle N^k(t) \rangle$ vs $\log \langle N^j (t) \rangle$ for $y = 1,2,\dots,Y$ (where $Y$ is the total number of available years for the selected species and $\langle N^k(t) \rangle$ is the $k$-th spatial sample moment in year $t$) gave the estimate of the sample exponent $b_{jk}$ for the selected species (Fig. 3). Frequency histograms of empirical exponents $b_{jk}$ are shown in Fig. \ref{histogram_bjk} (see also the box-whisker plots in the upper insets of Fig. 3); for every integer choice of $j$ and $k$ (here, up to $k=4$), the histogram is centered in $k/j$, as  the asymptotic model predicted. A one-sample $t$-test does not reject the null hypothesis that the sample mean of the values of $b_{jk}$ does not differ significantly from the theoretically predicted mean $k/j$ (Fig. \ref{histogram_bjk}).

\subsubsection{Inter-specific TL data analysis}
The inter-specific form of TL and the generalized scaling relationship for statistical moments (Eq. 2) were investigated following previous studies on conventional TL \cite{hanski82}. Spatial sample mean and variance were computed with the same dataset across similar sites. Data from sites labeled as B, C, X, AE in \cite{denboer77}, collected between 1961 and 1966, were used to calculate spatial moments across woodland sites. Data from sites labeled as AT, N, Z, AG in \cite{denboer77}, collected between 1963 and 1966, were used to calculate spatial moments across heath sites. As for the intra-specific TL analysis, we have limited the analysis of the inter-specific TL to the set of species that were present in all sites in each given year. Spatial moments of carabid beetles abundance were computed for each species individually and separately for each year and site type (woodland or heath). For each year, we calculated the least-squares slope of $\log \langle N^k \rangle$ versus $\log \langle N \rangle$ across all species at a given year and site type. Tables \ref{jk_inter-specific_table_wood}, \ref{jk_inter-specific_table_heath},  show the summary statistics for all years and site types. Figs. 4\textbf{(A)}, \ref{jk_inter-specific_all_years}\textbf{(M-N)} show the scaling of the $k$-th sample moment $\langle N^k \rangle$ with $\langle N \rangle$ when data for all years and site types are plotted together; each data point in Figs. 4, \ref{jk_inter-specific_all_years} refers to the spatial moments of a single species in one year and site type. Fig. \ref{jk_inter-specific_all_years}(\textbf{A-L}) shows the scaling of the $k$-th sample moment $\langle N^k \rangle$ with $\langle N \rangle$ for each year and site type separately. The least-squares exponents $b_{jk}$ computed in the linear regression of $\log \langle N^k \rangle$ versus $\log \langle N^j \rangle$ are compatible with the asymptotic model prediction $b_{jk}=k/j$ (Tables  \ref{jk_inter-specific_table_wood}, \ref{jk_inter-specific_table_heath}), as are the mean exponents $b_{jk}$ (Table \ref{mean_b_jk}).

\subsubsection{Test of the multiplicative growth model assumptions}
The Black Rock Forest dataset \cite{cohen13b} is known to conform to the hypotheses of the Lewontin-Cohen model \cite{lewontin69}, which is a particular case of the multiplicative growth model considered in the main text. A detailed account of the hypothesis testing can be found in \cite{cohen13b}.

In this section, we test the multiplicative growth model hypothesis on the carabid beetles dataset. The carabid beetles dataset consists of abundance data of carabid beetles ranging from a minimum of 3 to a maximum of 6 sites and from a minimum of 4 to a maximum of 6 consecutive years, depending on the species. We computed the multiplicative factors $A(p,t)=N(p,t)/N(p,t-1)$ separately for each species, site $p$ and pair of consecutive years.

Here we test some of the assumptions of the multiplicative growth model, namely the independence and identical distribution of multiplicative factors over sites and over time. {Each test was performed separately for each species. The tests employed rely on assumptions, such as normality of data, which were tested prior to performing the hypothesis testing. We excluded  from such tests the species for which the test assumptions were not met.} Tables \ref{IDp} and \ref{IDt} report the percentage of species for which a $p$-value smaller than $0.05$ was returned, when testing for the identical distribution of multiplicative factors over sites  and time, respectively. The number of species employed in each test, {that is, the number of species that met the test assumptions}, is reported in the third column of Tables \ref{IDp} and \ref{IDt}. The first four tests in Tables \ref{IDp} and \ref{IDt} test for identical mean and the last four tests test for identical variance. All tests were performed with the software \textit{Mathematica} v9.0.

The high percentages of rejection of the null hypotheses of equal mean and equal variance of multiplicative factors over sites and time in the carabid beetles dataset suggest that the carabid beetles population dynamics does not conform to the Markovian multiplicative growth model. Nevertheless, the predictions of the analysis regarding the higher-order sample exponents of the generalized TL were substantially confirmed. That the generalized TL pattern holds in the carabid beetles dataset, despite the disagreement with the assumptions of the Markovian multiplicative model, suggests that the results of our theoretical investigation might hold far beyond the population growth model considered in the main text.

\section{Supplementary text}
\subsection{Comparison with other demographic models}
The multiplicative growth model is one of numerous demographic models that predict TL. The exponent $b=2$ for the scaling of the variance versus the mean is typical of deterministic dynamics. For example, an exponential model of clonal growth \cite{cohen13a}, where clones grow exponentially with different growth rates (variability enters here only through the different growth rates and initial densities), and the above symmetric model for $\lambda=0$ or $\lambda=1$ both predict TL with exponent $2$. Although found in deterministic models, the exponent $b=2$ is also observed in stochastic models such as the continuous-time birth-death process and  the Galton-Watson branching process \cite{cohen14b}. Such models yield population exponents $b=2$ and $b=1$ respectively for asymptotically growing and decaying populations \cite{cohen14b}.

The theoretical investigation of multiplicative population processes showed that the generalized TL sample exponents $b_{jk}$ satisfy $b_{jk}\simeq k/j$ asymptotically for large $t$ for a broad ensemble of transition matrices $\Pi$ and sets of positive multiplicative factors. Additionally, our large-deviation approach and our small-sample argument suggest that the entropic term in Eq. \ref{exponent_finite} dominates over the other terms that contain the specifications of the demographic process. Thus the result might be more general than the class of multiplicative population growth models. We show here that $b_{jk}=k/j$ holds for the population exponents of other population growth processes, such as the birth-death process in the case of expanding populations.

{The moments of the birth-death process with constant birth rate $\lambda$ and constant death rate $\mu$ can be computed via the associated moment generating function $M$, which is equal to \cite{bailey64}:
\begin{equation}
M(\theta, t)=\left( \frac{\mu v(\theta, t)-1}{\lambda v(\theta,t)-1} \right)^{N_0},
\end{equation}
where $v(\theta,t)=\frac{\left( e^\theta-1\right)e^{(\lambda-\mu)t}}{\lambda e^\theta-\mu}$ and $N_0$ is the initial population size. The $k$-th moment of population size can be computed as $\langle N^k \rangle = \frac{\partial^k M(\theta, t)}{\partial \theta^k} |_{\theta=0}$. Here, we assume $N_0=1$ (but the result holds for any $N_0$) and an expanding population, i.e., $\lambda-\mu>0$. Because $v(0,t)=0$, $\frac{\partial v}{\partial \theta} (\theta, t)=(\lambda-\mu)e^{(\lambda-\mu)t} \frac{e^h}{(-e^h \lambda+\mu)^2}\propto e^{(\lambda-\mu)t}$ and $\frac{\partial^k v}{\partial \theta^k}(\theta,t)\propto e^{(\lambda-\mu)t}$, the leading term in the partial derivatives of $M(\theta,t)$ with respect to $\theta$, evaluated in $\theta=0$, can be written as:
\begin{equation}
\begin{aligned}
\frac{\partial^k M}{\partial \theta^k}(\theta,t) \biggr\rvert_{\theta=0} &= (-1)^{k+1}(\lambda-\mu) \lambda^{k-1} \frac{\left(\frac{\partial v}{\partial \theta}\right)^k}{(-1+\lambda v)^{k+1}}\biggr\rvert_{\theta=0}+o\left[\left(\frac{\partial v}{\partial \theta}\right)^k \biggr\rvert_{\theta=0} \right]\\
&= (\lambda-\mu)^{1-k}\lambda^{k-1} e^{k(\lambda-\mu)t}+o\left[e^{k(\lambda-\mu)t}\right],
\label{dkM}
\end{aligned}
\end{equation}
where the little-$o$ notation indicates that the remaining terms are negligible in the limit $t \to \infty$. {Derivation of the equation for $\frac{\partial^k M}{\partial \theta^k}(\theta,t)$ (first line of Eq. \ref{dkM}) shows that the leading term in $\frac{\partial^{k+1} M}{\partial \theta^{k+1}}(\theta,t) \biggr\rvert_{\theta=0}$ is equal to $(\lambda-\mu)^{k}\lambda^{k} e^{(k+1)(\lambda-\mu)t}+o\left[e^{(k+1)(\lambda-\mu)t}\right]$, which coincides with replacing $k$ by $k+1$ in Eq. \ref{dkM}}. Eq. \ref{dkM} can be obtained by considering that, because $\frac{\partial^k v}{\partial \theta^k}\propto e^{(\lambda-\mu)t}$ and $v(0,t)=0$, the leading term in $\frac{\partial M}{\partial \theta}(\theta,t) = (\lambda-\mu) \frac{{\partial v}/{\partial \theta}}{(-1+\lambda v)^2}$ evaluated at $\theta=0$ is the second term in the quotient rule $(f/g)'=(f'g-fg')/g^2$, that is, the term that raises the exponent of $\frac{\partial v}{\partial \theta}$ by one unit. For subsequent derivatives, the quotient rule is applied to the leading term. All other terms in $\frac{\partial^k M}{\partial \theta^k}(\theta,t) \big\rvert_{\theta=0}$ contain products of partial derivatives\footnote{For example:
\begin{align}
\frac{\partial^2 M}{\partial \theta^2} \biggr\rvert_{\theta=0} &= (\lambda-\mu) \left( 2 \lambda \frac{\partial v}{\partial \theta} \biggr\rvert_{\theta=0}^2 + \frac{\partial^2 v}{\partial \theta^2} \biggr\rvert_{\theta=0} \right),\\
\frac{\partial^3 M}{\partial \theta^3} \biggr\rvert_{\theta=0} &= (\lambda-\mu) \left( 6 \lambda^2 \frac{\partial v}{\partial \theta} \biggr\rvert_{\theta=0}^3 + 6 \lambda  \frac{\partial v}{\partial \theta} \frac{\partial^2 v}{\partial \theta^2} \biggr\rvert_{\theta=0} + \frac{\partial^3 v}{\partial \theta^3} \biggr\rvert_{\theta=0} \right).
\end{align}
}, i.e., $\prod_{j=1}^k ( \frac{\partial^{j} v}{\partial \theta^{j}} )^{q_j}$, with $\sum_{j=1}^k q_j < k$ (with $q_j \in \mathbb{N}$) and are thus negligible in the limit $t \to \infty$. From Eq. \ref{dkM} it follows that $\lim_{t \to \infty} \frac 1t \log \langle N^k \rangle = k (\lambda-\mu)$; thus, the generalized TL holds with $b_{jk}=k/j$.}

{The asymptotic behavior of exponents, i.e., $\lim_{t \to \infty} \frac 1t \log \langle N^k \rangle = k (\lambda-\mu)$, can also be computed via the continuous approximation of the birth-death process. Although such calculations do not provide further understanding of the birth-death process (we have already calculated the limiting behavior of $\langle N^k \rangle$ for large $t$), the fact that the continuous approximation of the birth-death process coincides with that of the Galton-Watson branching process \cite{feller57,harris63,rubin14} suggests an even broader validity for the generalized TL result $b_{jk}=k/j$. The detailed calculation of exponents in the continuous approximation of the birth-death process and the Galton-Watson branching process is provided in the following section.
\subsubsection{Moments of population density in the continuous approximation of the birth-death process and the Galton-Watson branching process}
The forward Kolmogorov equation for the continuous approximation of the birth-death process reads \cite{feller57,harris63,rubin14}:
\begin{equation}
\frac{\partial p(x,t)}{\partial t}=- \alpha \frac{\partial [ x p(x,t) ]}{\partial x}+\frac \beta2 \frac{\partial^2 [x p(x,t)]}{\partial x^2},
\label{fkolmogorov}
\end{equation}
where $p(x,t)$ is the probability density function for the population density $x$ at time $t$ (here, $x\in \mathbb{R}$ is the population density and should not be confused with the frequency of multiplicative factors used in the rest of the paper). Eq. \ref{fkolmogorov} is the continuous approximation of a birth-death process with birth rate $\lambda$ and death rate $\mu$ such that $\alpha=\lambda - \mu$ and $\beta=\lambda+\mu$. Eq. \ref{fkolmogorov} also arises as the continuous approximation of the Galton-Watson branching process for large populations \cite{feller57,harris63,rubin14}. The solution of Eq. \ref{fkolmogorov} with initial condition $x(0)=x_0$ is known \cite{bailey64} and is equal to:
\begin{equation}
p(x,t)=\frac{2 \alpha}{\beta(e^{\alpha t}-1)}\left( \frac{x_0 e^{\alpha t}}{x} \right)^{\frac 12} \exp \left[ \frac{-2\alpha (x_0 e^{\alpha t}+x)}{\beta(e^{\alpha t}-1)} \right] I_1\left[ \frac{4 \alpha(x_0 x e^{\alpha t})^{\frac 12}}{\beta(e^{\alpha t}-1)} \right],
\label{solution}
\end{equation}
where $I_1$ is the modified Bessel function of the first kind. Differentiation with respect to $\gamma$ of the identity $\int_0^\infty dx I_1(x) e^{-\gamma x^2}=e^{1/(4\gamma)}-1$  gives the following equation:
\begin{equation}
C \int_0^\infty dx x^k x^{-\frac 12} I_1(x^{\frac 12} A) e^{Bx} = 2 C A^{-(2k+1)}\left(-\frac{d}{d\gamma}\right)^k \biggr\rvert_{\gamma=-\frac{B}{A^2}}\left( e^{\frac 1{4\gamma}}-1\right),
\end{equation}
which allows calculating the moments of Eq. \ref{solution} with $A=\frac{4 \alpha \left( x_0 e^{\alpha t} \right)^{\frac 12}}{\beta \left( e^{\alpha t}-1 \right)}$, $B=\frac{2\alpha}{\beta \left( e^{\alpha t}-1 \right)}$ and $C=\frac{2 \alpha \left( x_0 e^{\alpha t} \right)^{\frac 12}}{\beta(e^{\alpha t}-1)} \exp \left[ -\frac{2\alpha x_0 e^{\alpha t}}{\beta \left( e^{\alpha t}-1 \right)} \right]$. For an expanding population, $\alpha>0$; thus asymptotically for large $t$:
\begin{equation}
\begin{aligned}
A &\propto e^{-\frac{\alpha t}2},\\
B &\propto e^{-\alpha t},\\
C &\propto e^{-\frac{\alpha t}2}.
\end{aligned}
\end{equation}
Therefore, $\gamma=-\frac{B}{A^2}$ tends to a constant and one has:
\begin{equation}
\langle x^k \rangle \propto C A^{-2k+1} \propto \left( e^{\alpha t} \right)^k,
\end{equation}
which implies that, asymptotically, the generalized TL holds with exponent $b_{jk}=k/j$.}


\section{Supplementary figures}
\begin{figure}[h]
\begin{center}
\includegraphics[width=7.5cm]{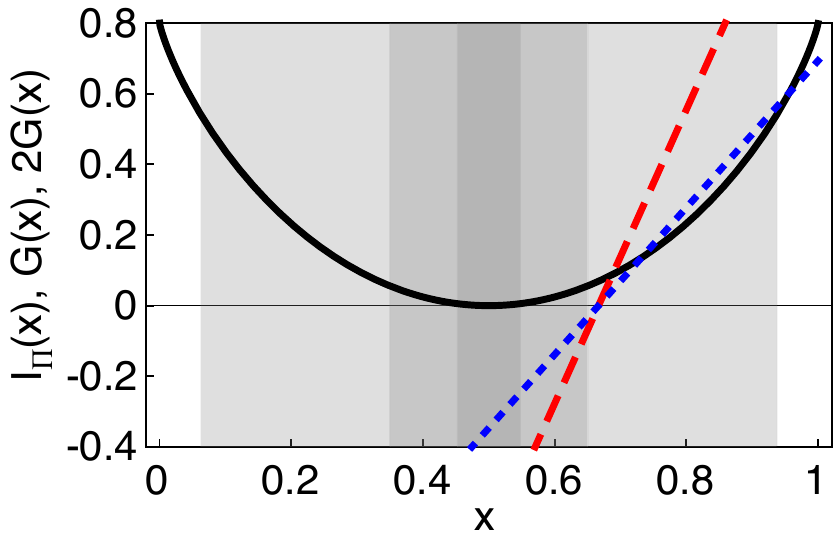}
\caption{Plot of $I_{\Pi}(x)$ (black curve), $G(x)$ (dotted blue line) and $2G(x)$ (dashed red line). Marked in gray are the regions $[x_-, x_+]$ at times $t=10$, $100$ and $1000$ (from light to dark gray) for fixed $R=100$. These gray regions are the intervals over which the supremum in Eq. \ref{exponent_finite} is computed. In this example, $r=2$, $s=1/4$, $\lambda=0.55$. The quantities $x_+$ and $x_-$ are computed by solving numerically Eq. \ref{exponential_R}.}
\label{figI_G}
\end{center}
\end{figure}
\begin{figure}[h]
\begin{center}
\includegraphics[width=7.5cm]{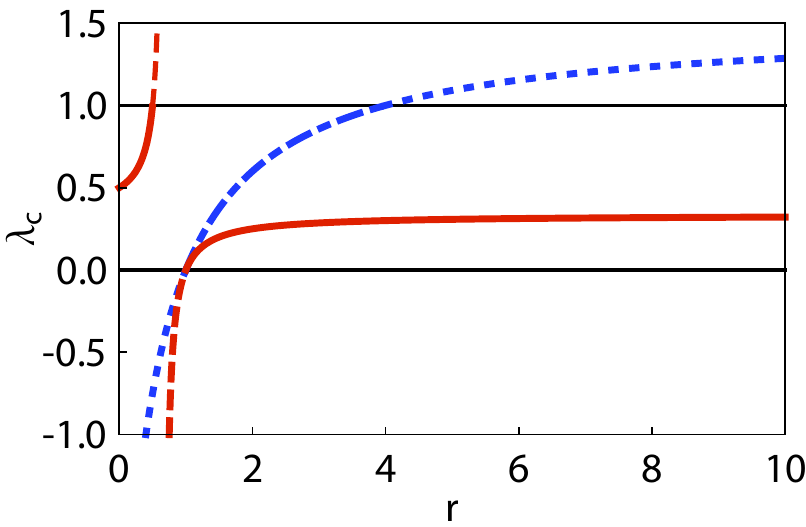}
\caption{The critical transition probability $\lambda_c$ as a function of $r$ (with $s$ fixed). Below the black horizontal line at $\lambda_c=0$ and above the black horizontal line at $\lambda_c=1$, $\lambda_c$ does not exist. The red (solid for $0\leq \lambda_c \leq 1$ and dashed otherwise) and blue (dash-dotted for $0 \leq \lambda_c \leq 1$ and dotted otherwise) lines $\lambda_c=\frac{1-r-s+rs}{-r-s+2 rs}$ were calculated by solving Eq. \ref{critical} with respect to $\lambda_c$ with, respectively, $s=2$ and $s=1/4$. For any given $s$, $\lambda_c=0$ for $r=1$ and $\lambda_c=1$ for $r=1/s$.}
\label{figure11}
\end{center}
\end{figure}
\begin{figure}[p]
\begin{center}
\includegraphics[width=0.9\columnwidth]{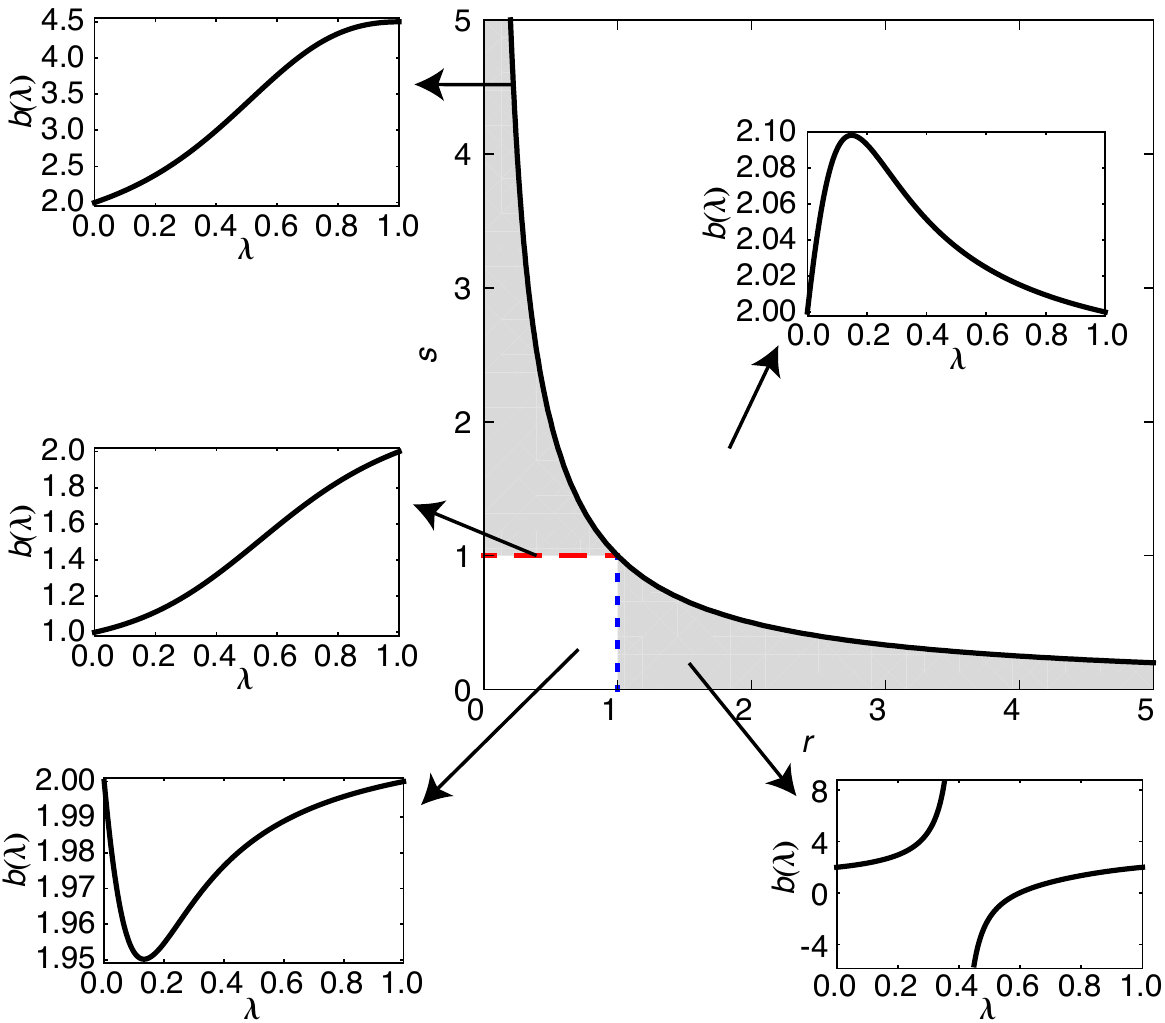}
\caption{Existence of a critical transition probability $\lambda_c$. Smaller panels show the population exponent $b(\lambda)$ (Eq. \ref{exponent_infinite}) for various choices of the multiplicative factors in different regions of the plane $(r,s)$ (larger panel).  Only in the interior of the gray region of the plane $(r,s)$, $\lambda_c$ exists. The solid black line represents the curve $rs=1$.}\label{b_rs}
\label{phase}
\end{center}
\end{figure}
\begin{figure}[p]
\begin{center}
\includegraphics[width=0.6\columnwidth]{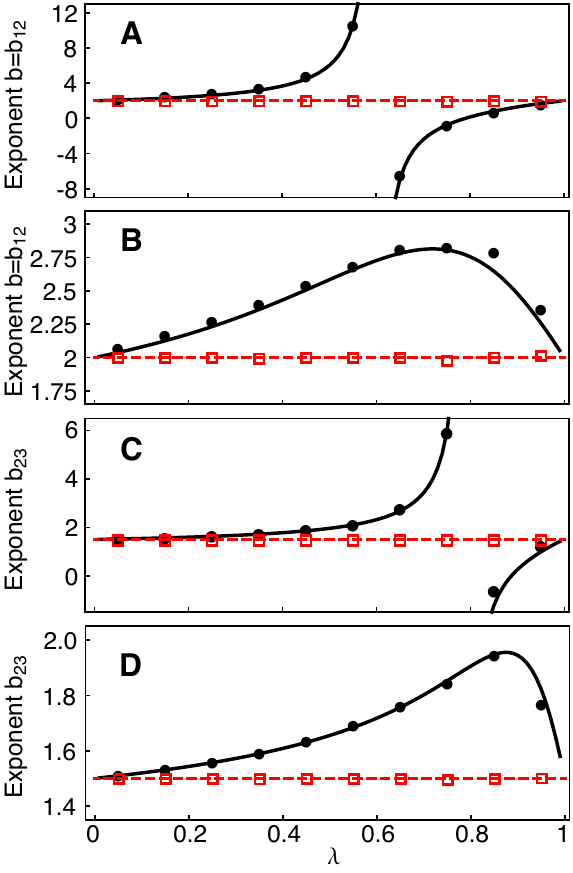}
\caption{TL exponent $b=b_{12}$ and generalized exponent $b_{23}$ for different values of the transition probability $\lambda$ (panels \textbf{A-B} as in Fig. 1). The sample exponents computed in simulations of a $2$-state multiplicative process with symmetric transition matrix in the two regimes $1 \ll t \ll \log R$ (black filled dots, $R=10^6$ up to time $t=10$) and $t \gg \log R$ (red open squares, $R=10^4$ up to time $t=400$) are in good agreement with predictions for the asymptotic population (black solid line, Eq. S6) and sample (red dashed line, $b=b_{12}=2$ and $b_{23}=3/2$) exponents. In the simulations, the sample exponent $b=b_{12}$ was computed by least-squares fitting of $\log \mbox{Var}[N(t)]$ as a function of $\log \mathbb{E}[N(t)]$ for the last $6$ (black dots) and $200$ (red squares) time steps. The sample exponent $b_{23}$ was computed by least-squares fitting of $\log \mathbb{E}[N(t)^3]$ as a function of $\log \mathbb{E}[N(t)^2]$ in the same fashion. In panel \textbf{(A)}, which reproduces \cite{cohen14a}, and \textbf{(C)} $\chi=\{r,s\}=\{2,1/4\}$ ($b=b_{12}$ and $b_{23}$ display discontinuities); in panel \textbf{(B)} and \textbf{(D)} $\chi=\{r,s\}=\{4,1/2\}$ (in such a case, $b_{12}$ and $b_{23}$ display no discontinuities).}
\label{b}
\end{center}
\end{figure}
\begin{figure}[p]
\begin{center}
\includegraphics[width=\columnwidth]{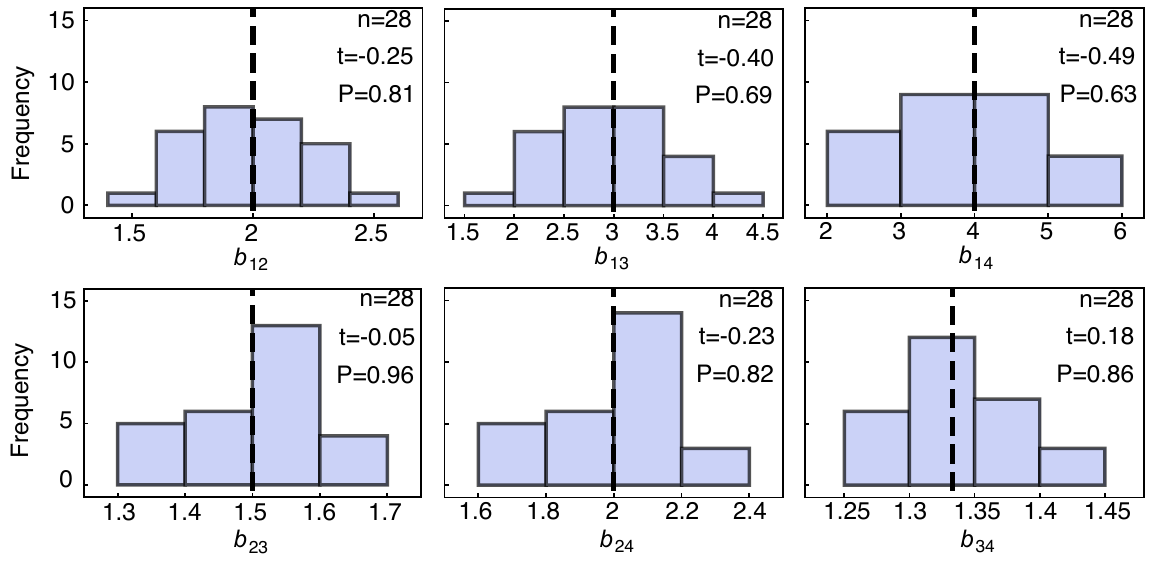}
\caption{Frequency histogram for the exponent $b_{jk}$ in the intra-specific generalized TL $\langle N^k \rangle = a \langle N^j \rangle^{b_{jk}}$, computed for each species (carabid beetles, \cite{denboer77}) across similar sites (woodland or heath). The dashed black line shows the value of the exponent $b_{jk}=k/j$ as the asymptotic model predicted. The binning of data points is determined by using Scott's rule \cite{scott79}. Shown in each panel are the number of observations $n$ of $b_{jk}$, the test statistic for the $t$-test of the null hypothesis that the sample mean of the values of $b_{jk}$ did not differ significantly from the theoretically predicted mean $k/j$ and the corresponding $p$-value.}
\label{histogram_bjk}
\end{center}
\end{figure}
\clearpage
\thispagestyle{empty}
\begin{figure}[p]
\begin{center}
\includegraphics[width=0.85\columnwidth]{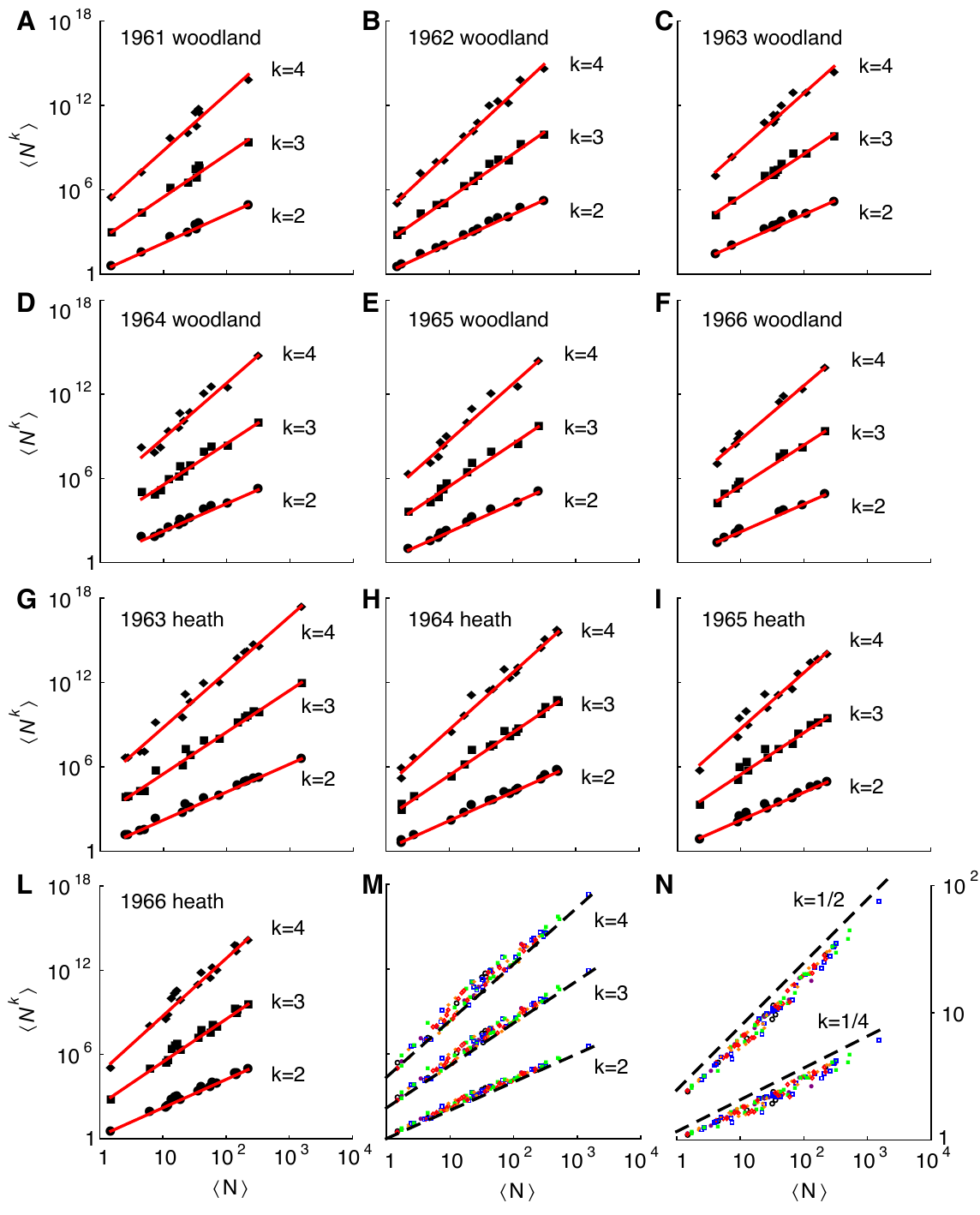}
\caption{Generalized TL for inter-specific patterns of abundance of carabid beetles, data from \cite{denboer77}. \textbf{(A-L)} Double logarithmic plots of $\langle N^k \rangle$ vs $\langle N \rangle$ for all species in separate years and site type (black symbols). The red lines show the least squares regressions of $\log \langle N^k \rangle$ vs $\log \langle N \rangle$ (Tables \ref{jk_inter-specific_table_wood} and \ref{jk_inter-specific_table_heath}). Offsets are introduced in the data and in the linear regressions to aid visual inspection. \textbf{(M-N)} Double logarithmic plot of $\langle N^k \rangle$ vs $\langle N \rangle$ for all species, years and site type, with integer \textbf{(M)} and non-integer \textbf{(N)} $k$. Each data point refers to sample moments computed for a single species in one year and site type. The color and symbol code identifies data relative to the same year: 1961 (black open circles), 1962 (purple filled circles), 1963 (blue open squares), 1964 (green filled squares), 1965 (orange filled diamonds), 1966 (red open diamonds). The color and symbol code does not distinguish site type. Dashed black lines of slope $b_{1k}=k/1=k$ (asymptotic model prediction for the sample exponent) and arbitrary intercept are shown in each plot. Offsets are introduced in the data to aid visual inspection. }
\label{jk_inter-specific_all_years}
\end{center}
\end{figure}
%
%

\clearpage
\section*{Supplementary tables}
\begin{table}[htdp]
\caption{Sample exponents for the generalized TL in the Black Rock Forest dataset, data from \cite{cohen13b}.}
\begin{center}
\begin{tabular}{|c|c|c|c|}\hline
(j,k)	&	$k/j$	&	$b_{jk} \pm $SE &	$R^2$	\\ \hline
1,2 	&	$2$	&	$2.14 \pm 0.12$	&	0.991\\
1,3		&	$3$	&	$3.33 \pm 0.32$	&	0.973\\
1,4		&	$4$	&	$4.54 \pm 0.58$	&	0.954\\
2,4		&	$2$	&	$2.15 \pm 0.16$	&	0.984\\
2,3		&	$1.5$	&	$1.57 \pm 0.07$	&	0.995\\
3,4		&	$1.333$	&	$1.37 \pm 0.04$	&	0.997\\
1,1/2	&	$0.5$	&	$0.48 \pm 0.02$	&	0.997\\
1,1/4	&	$0.25$	&	$0.23 \pm 0.01$	&	0.993\\
1,2/3	&	$0.667$	&	$0.65 \pm 0.01$	&	0.999 \\ \hline
\end{tabular}
\end{center}
\label{table_BRF}
\end{table}
\begin{table}[!h]
\caption{Sample exponents for the inter-specific generalized TL on carabid beetles abundances in woodland sites, data from \cite{denboer77}. The column $k/j$ gives the asymptotic model prediction for the exponent $b_{jk}$. The estimates $b_{jk}$ (mean$\pm$SE) are the least-squares slopes of $\log \langle N^k \rangle$ vs $\log \langle N \rangle$. $R^2$ is the squared correlation coefficient. Nonlinearity was checked with least-squares quadratic regression on log-log coordinates. The coefficient of the second power term did not differ significantly from $0$ in any of the regressions; hence, the null hypothesis of linearity was not rejected.}
\begin{center}
\begin{tabular}{|c|c|c|c|c|c|c|c|}
\multicolumn{2}{c}{} & \multicolumn{2}{c}{1961} & \multicolumn{2}{c}{1962}	& \multicolumn{2}{c}{1963} \\ \hline
$j,k$	&	$k/j$	&	$b_{jk} \pm$SE	& $R^2$	&	$b_{jk} \pm$SE	& $R^2$	&	$b_{jk} \pm$SE	& $R^2$		\\ \hline
1,2	& $2$	&	$2.03\pm0.09$	&	$0.988$	&	$2.07\pm0.04$	&	$0.995$	&	$2.00\pm0.07$	&	$0.988$	\\
1,3	& $3$	&		$3.04\pm0.18$	&	$0.976$	&	$3.13\pm0.09$	&	$0.991$	&	$3.00\pm0.15$	&	$0.977$		\\
1,4	& $4$	&		$4.03\pm0.28$	&	$0.968$	&	$4.20\pm0.14$	&	$0.988$	&	$4.01\pm0.23$	&	$0.971$		\\ \hline
\multicolumn{2}{|c|}{No. points} & \multicolumn{2}{|c|}{$9$} & \multicolumn{2}{|c|}{$13$}	& \multicolumn{2}{|c|}{$11$} \\ \hline
\multicolumn{8}{c}{}\\
\multicolumn{2}{c}{} & \multicolumn{2}{c}{1964} & \multicolumn{2}{c}{1965}	& \multicolumn{2}{c}{1966} \\ \hline
$j,k$	&	$k/j$	&	$b_{jk} \pm$SE	& $R^2$	&	$b_{jk} \pm$SE	& $R^2$	&	$b_{jk} \pm$SE	& $R^2$		\\ \hline
1,2	& $2$	&	$1.96\pm0.09$	&	$0.977$	&	$2.01\pm0.07$	&	$0.989$	&	$1.97\pm0.06$	&	$0.995$	\\
1,3	& $3$	&		$2.94\pm0.20$	&	$0.957$	&	$3.00\pm0.16$	&	$0.976$	&	$2.90\pm0.12$	&	$0.989$		\\
1,4	& $4$	&		$3.92\pm0.29$	&	$0.947$	&	$4.00\pm0.24$	&	$0.967$	&	$3.83\pm0.18$	&	$0.985$		\\ \hline
\multicolumn{2}{|c|}{No. points} & \multicolumn{2}{|c|}{$12$} & \multicolumn{2}{|c|}{$11$}	& \multicolumn{2}{|c|}{$9$} \\ \hline
\end{tabular}
\end{center}
\label{jk_inter-specific_table_wood}
\end{table}
\begin{table}[htdp]
\caption{Sample exponents for the inter-specific generalized TL on carabid beetles abundances in heath sites, data from \cite{denboer77}. The Table is organized as Table \ref{jk_inter-specific_table_wood}.}
\begin{center}
\begin{tabular}{|c|c|c|c|c|c|}
\multicolumn{2}{c}{} & \multicolumn{2}{c}{1963} & \multicolumn{2}{c}{1964} \\ \hline
$j,k$	&	$k/j$	&	$b_{jk} \pm$SE	& $R^2$	&	$b_{jk} \pm$SE	& $R^2$		\\ \hline
1,2	& $2$	&	$1.99\pm0.05$	&	$0.993$	&	$2.02\pm0.04$	&	$0.995$	\\
1,3	& $3$	&		$2.98\pm0.09$	&	$0.987$	&	$3.03\pm0.08$	&	$0.990$	\\
1,4	& $4$	&		$3.83\pm0.18$	&	$0.985$	&	$3.96\pm0.14$	&	$0.983$	\\	\hline
\multicolumn{2}{|c|}{No. points} & \multicolumn{2}{|c|}{$16$} & \multicolumn{2}{|c|}{$16$}	\\ \hline
\multicolumn{6}{c}{}\\
\multicolumn{2}{c}{} & \multicolumn{2}{c}{1965} & \multicolumn{2}{c}{1966} \\ \hline
$j,k$	&	$k/j$	&	$b_{jk} \pm$SE	& $R^2$	&	$b_{jk} \pm$SE	& $R^2$		\\ \hline
1,2	& $2$	&	$1.98\pm0.08$	&	$0.982$		&	$2.02\pm0.06$	&	$0.986$	\\
1,3	& $3$	&	$2.97\pm0.17$	&	$0.965$		&	$3.05\pm0.13$	&	$0.974$		\\
1,4	& $4$	&	$4.04\pm0.12$	&	$0.987$		&	$3.98\pm0.26$	&	$0.956$	\\ \hline
\multicolumn{2}{|c|}{No. points} &	\multicolumn{2}{|c|}{$13$}& \multicolumn{2}{|c|}{$17$} \\ \hline
\end{tabular}
\end{center}
\label{jk_inter-specific_table_heath}
\end{table}
\begin{table}[htdp]
\caption{Statistics of estimated sample exponents in the inter-specific generalized TL. The column $k/j$ gives the asymptotic model prediction for the exponent $b_{jk}$. The point estimate is computed as the average $b_{jk}$ across years and site type, not by pooling all the data from different years and site types to calculate means and variances. The confidence intervals are obtained via bootstrapping with $10^6$ bootstrap samples from the set of $b_{j,k}$.}
\begin{center}
\begin{tabular}{|c|c|c|c|c|} \hline
$j,k$	&	$k/j$	&	$b_{jk}$ point estimate	& 	$2.5\%$ percentile	& 	$97.5\%$ percentile	\\ \hline
1,2	& $2$	&	$2.005$	&	$1.984$	&	$2.025$	\\
1,3	& $3$	&		$3.005$	&	$2.961$	&	$3.042$	\\
1,4	& $4$	&		$3.994$	&	$3.936$	&	$4.057$	\\ \hline
\end{tabular}
\end{center}
\label{mean_b_jk}
\end{table}
\begin{table}[htdp]
\caption{Tests of whether  multiplicative growth factors of carabid beetle abundances have the same means and variances over sites. The table reports the percentage of $p$-values smaller than $0.05$ across all species, for several statistical tests. The percentage refers to the number of species used in the test, reported in the third column.}
\centering
\begin{tabular}{|l|c|c|}\hline
Test	&	$\%$ of $p<0.05$		&	N. species \\ \hline
Complete	Block F	&	4.3\%	&	23 \\
Friedman Rank		&	4.2\%	&	24 \\
Kruskal Wallis		&	0\%		&	24 \\
K Sample T		&	0\%		&	23 \\ \hline
Bartlett 			&	29.6\%	&	27 \\
BrownForsythe		&	3.7\%	&	27 \\
Conover			&	7.1\%	&	28 \\
Levene			&	25.9\%	&	27 \\ \hline
\end{tabular}
\label{IDp}
\end{table}
\begin{table}[htdp]
\caption{Tests of whether  multiplicative growth factors of carabid beetle abundances have the same means and variances over years. The table reports the percentage of $p$-values smaller than $0.05$ across all species for several statistical tests. The percentage refers to the number of species used in the test, reported in the third column.}
\centering
\begin{tabular}{|l|c|c|}\hline
Test	&	$\%$ of $p<0.05$	&	N. species\\ \hline
Complete	Block F 	&	14.3\%	&	14	\\
Friedman Rank		&	20.0\%	&	15	\\
Kruskal Wallis		&	53.3\%	&	15	\\
K Sample T		&	35.7\%	&	14	\\	\hline
Bartlett 			&	48.1\% 	&	27	\\
BrownForsythe		&	7.4\%	&	27 	\\
Conover			&	14.3\%	&	28	\\
Levene			&	51.9\%	&	27	\\ \hline
\end{tabular}
\label{IDt}
\end{table}

\clearpage


\begin{thebibliography}{10}
\bibitem{logR}
 More precisely, the sample exponent is close to the population exponent if the
  suprema in the equation for the population exponent $b_{jk}$
  \cite{derivazione} are included in $[x_-, x_+]$. Using $x_\pm\simeq \frac 12
  \pm \sqrt{ \frac{1-\lambda}{2\lambda} \frac{\log R}t }$, if the largest
  supremum is $x_{\max}>1/2$, the sample exponent approximates accurately the
  population exponent for $t < \frac{1-\lambda}{\lambda}(x_{\max}-1/2)^{-2}\log
  R$ and tends to $k/j$ for $t >
  \frac{1-\lambda}{\lambda}(x_{\max}-1/2)^{-2}\log R$ (Figs. 1,2).
\end{thebibliography}

%
\end{document}